# Control of turn-to-turn contact resistivity in resistively insulated REBCO coils

Jun Lu, Kwangmin Kim, Iain Dixon, Justin Deterding, Emsley Marks, Brent Jarvis,

Denis Markiewicz, Hongyu Bai, and Mark Bird

National High Magnetic Field Laboratory, Tallahassee, FL 32310

**Abstract**

Resistively insulated (RI) REBCO magnets feature short ramp times and low ramp losses while maintaining the advantages of no-insulation coils with high engineering current density and tolerance for defects in the REBCO conductor. Control of the turn-to-turn contact resistivity ($\rho_c$) is key to RI technology. $\rho_c$ must be sufficiently high to prevent a large transient current, which could result in high mechanical stress during magnet quenches. Meanwhile it must be lower than the quench propagation limit to avoid conductor burn-out during a quench. Therefore, it is critical to control $\rho_c$ within a suitable range of values which is usually coil specific.

Previously, we discovered that $\rho_c$ between two REBCO tapes with a stainless steel interlayer decreases dramatically with contact pressure cycling by up to three orders of magnitude. This drastic change made it impossible to design a suitable $\rho_c$ value for a stainless steel co-wound RI magnet. In this work, we first present methods for mitigating $\rho_c$ pressure cycling sensitivity. We found that by adding conductive fillers, such as conductive paste or epoxy, the $\rho_c$ load cycling sensitivity is largely mitigated. For dry-wound coils, $\rho_c$ load cycling sensitivity is mitigated by coating REBCO tape with a layer of 2- 3 $\mu$m of PbSn solder. In addition, $\rho_c$ can be controlled by oxidizing the stainless steel co-wind tape by heating stainless steel tapes at different temperatures in air. Using above methods, short sample tests showed that $\rho_c$ was controlled to prescribed values of 1000 and 5000 $\mu\Omega$-cm$^2$ and was not sensitive to contact pressure cycling up to 30,000 cycles at 4.2 K. The new $\rho_c$ control method was applied to a 6 double-pancake test coil which was tested at 4.2 K. The $\rho_c$ in this test coil was comparable with the short sample results. This demonstrated the ability of this new method to control $\rho_c$ in large coils. Moreover, we propose a method of measuring coil's $\rho_c$ by the decay time of coil's inductive voltage.

1. Introduction

No-insulation (NI) and resistive-insulation (RI) REBCO magnet technologies allow for a high engineering current density compared with their conventional insulation counterparts. They have been developing rapidly over the past 15 years. Using NI technology, an extremely high field of 45.2 T has been achieved in a background field of 31 T [1]. In addition, fields over 30 T had been achieved by NI/RI REBCO insert coils in the background field provided by either superconducting or resistive magnets [2] - [5]. The key to NI/RI technology is the control of turn-to-turn contact resistivity ($\rho_c$). A low $\rho_c$ in NI magnets results in long charging delays and coupling loss between turns during field ramps, which are undesirable for a user magnet that requires frequent field ramps. In addition, during a magnet quench low $\rho_c$ results in large transient current in both radial and azimuthal directions [5], [6] which in turn causes high mechanical stresses that may be difficult to mitigate. On the other hand, a high $\rho_c$ compromises the advantage of stability and quench self-protection ability and risks burn-outs at hotspots in the magnet. For these reasons, there have been many studies on $\rho_c$ measurements and control in recent literature. The studies on $\rho_c$ measurements including the ones by our group can be found in [7] – [17]. In the afore mentioned NI/RI magnets that reached 30 T and above, none of them had conductor or co-wind surface engineered to achieve a prescribed $\rho_c$. To develop methods to control $\rho_c$, several groups have achieved various levels of success [18] - [24]. However, a reliable control method that can engineer a prescribed $\rho_c$ value in a wide range and is proved in large RI coil tests is still lacking.

Since high-field REBCO magnets are often stress-limited, co-wound stainless steel tapes are often used for mechanical reinforcement [25, 26]. This allows, in principle, to implement resistive-insulation techniques by controlling the contact resistance between REBCO and co-wind tape. For this reason, the contact resistance of interface between REBCO and stainless steel was studied [11]. We discovered however, that contact pressure cycling at 4.2 K dramatically reduces $\rho_c$ [7],[8],[11]. This considerable $\rho_c$ variation during the lifetime of a magnet makes control of $\rho_c$ practically impossible.

In this study, we developed a method to mitigate the load cycling sensitivity of $\rho_c$ by coating REBCO coated conductor with eutectic PbSn solder. Subsequently, the control of $\rho_c$ was achieved by surface oxidation of the stainless steel co-wind tape at different temperatures. This method was successfully used in a 6 double-pancake test coil named PTC-6. The details of the development of this method are described, and data from PTC-6 tests will be presented. As a part of PTC-6 coil testing, a new method of measuring coil $\rho_c$ was proposed and demonstrated.

## 2. Experiment

REBCO tapes were SCS4050-AP by SuperPower Inc. They were 4 mm wide with 50 μm thick Hastelloy C-276 substrate and 20 μm Cu stabilizer on each side. The stainless steel co-wind were 316L half hard tapes that were 4 mm wide and 25 μm thick, procured from Ulbrich of Illinois, Inc.

$\rho_c$ tests using short samples were performed at 77 and 4.2 K in a device as shown in Figure 1 [7], [8]. A REBCO/stainless steel/REBCO stack with a 25 mm long overlap was put under load in tape's thickness direction in liquid nitrogen or liquid helium bath. The load was cycled between 250 and 2500 N (corresponds to 2.5 – 25 MPa contact pressure) at 5 Hz by a hydraulic universal test machine (MTS corporation). The contact resistance was measured by a 4-probe method with a current of 1 A.

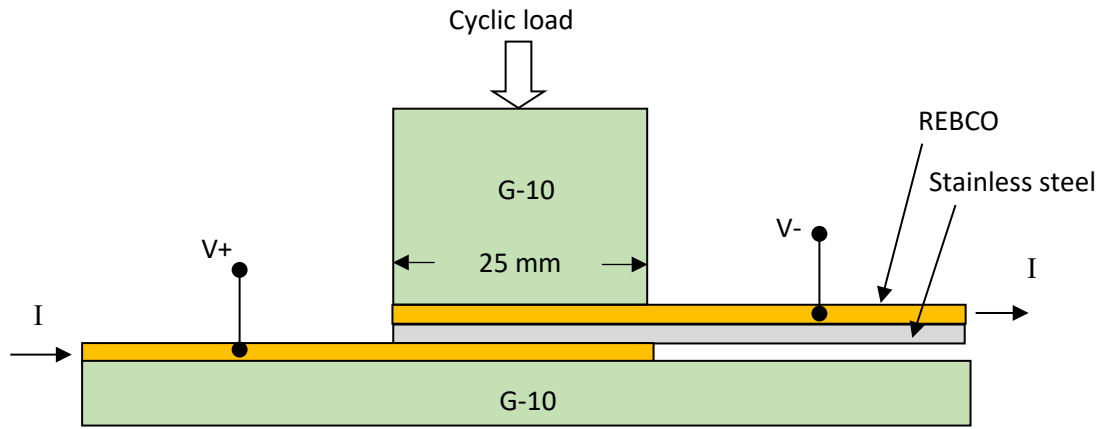

*Figure 1. Schematic of contact resistance test device. The device was placed in a universal test machine, and tests were performed in liquid helium or liquid nitrogen bath.*

A key step in this work was mitigation of load cycling sensitivity by coating a layer of relatively soft material on REBCO tapes. For this, we developed a reel-to-reel solder coating machine, a schematic of which is shown in Figure 2. The as-received REBCO tape went through a molten eutectic PbSn solder pot at 240 °C. The excessive molten solder on REBCO tap was then wiped by a pair of silicone pads as soon as it came out of the solder pot. In this process, the typical linear speed of REBCO tape was 6 m/min. The resultant coating thickness was about 2 - 3 µm as measured by cross-sectional microscopy. The effect of the coating process on REBCO properties was checked by measuring 77 K self-field Ic and RRR of the copper stabilizer before and after coating. No degradation was observed within the measurement error.

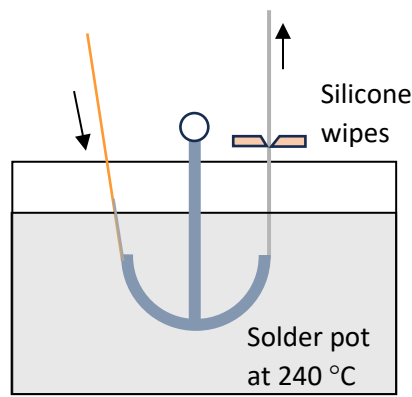

*Figure 2. Schematic of the solder coating process. REBCO tape slid through the half-circle fixture in the solder pot. The half-circle fixture has a radius of 40 mm. The solder was eutectic PbSn. The linear speed was 6 m/min.*

We control $\rho_c$ via surface treatment of the stainless steel co-wind tape. One method was oxidizing it in air in a vertical furnace. We adjust the furnace temperature to control the degree of oxidation, which in turn controls $\rho_c$. The linear speed of this reel-to-reel process was 1 m/min. Additional methods of oxidation were also explored. For instance, we immersed stainless steel samples in a solution of Ultra-Blak®407L (Electrochemical Products, Inc.) at 120 °C for different times. We also experimented with coating the 600 atomic layers of TiN and $Al_2O_3$ by atomic layer deposition.

To demonstrate the effectiveness of the $\rho_c$ control, an RI test coil, named as PTC-6, was designed fabricated and tested. The coil parameters of PTC-6 are listed in Table 1. It consists of 6 double-pancakes wound with solder-coated REBCO tapes and co-wound with surface treated stainless steel tapes. $a_1$, $a_2$, and $a_3$ are the inner, outer and overband radius respectively. The winding stress for both REBCO and stainless steel co-wind was 35 MPa. The targeted $\rho_c$ was 1000 $\mu\Omega$-cm$^2$ decided based on coil quench considerations.

Table 1 Parameters of PTC-6 test coil

| Parameters | PTC-6 |
| --- | --- |
| Number of double-pancakes | 6 |
| $a_1$ (mm) | 22 |
| $a_2$ (mm) | 40 |
| $a_3$ (mm) | 42 |
| Number of turns per disk | 123 |
| $I_{max}$ (A) | 240 |
| $B_{z-max}$ (T) | 11.3 |
| Inductance (mH) | 82.78 |
| Winding stress (MPa) | 35 |
| Targeted $\rho_c$ ($\mu\Omega$-cm$^2$) | 1000 |

The coil was tested at 4.2 K in a 6.5 T background field provided by a NbTi magnet. The average $\rho_c$ was determined by decay time of the central field after a sudden discharge at 5 A or by decay time of the inductive voltage after a ramp-and-hold operation. The decay time constant $\tau$ was determined by the time for the signal to decay to 1/e of its original value. The average contact resistance $R_c$ of the entire coil was calculated using a simple lumped circuit model,

$$R_c = L/\tau \qquad (1)$$

where *L* is the coil's inductance which was calculated and confirmed by the inductive voltage value during a current ramp. Subsequently, the contact resistivity $\rho_c$ was calculated by,

$$R_c = \sum_{i=1}^{N} \frac{\rho_c}{2\pi r_i w} = \frac{\rho_c}{2\pi w} \sum_{i=1}^{N} \frac{1}{r_i} \qquad (2)$$

$$\rho_c = 2\pi w R_c / \sum_{i=1}^{N} \frac{1}{r_i} \qquad (3)$$

where *N* is the total number of turns of all pancakes, and $r_i$ the radius of $i^{th}$ turn, *w* the tape width.

## 3. Results

### 3.1 Preliminary development for $\rho_c$ control

Control of $\rho_c$ is the ability to adjust $\rho_c$ to a specified value for RI magnet design. In case of REBCO/stainless steel contact, an obvious solution is to form a thin oxide layer on stainless steel with controlled thickness [11]. Figure 3 shows $\rho_c$ of REBCO/stainless steel/REBCO stack at 4.2 K as a function of load cycle. REBCO tapes were in as-received condition, while the stainless steel tapes were oxidized in air at different temperatures for 1 minute. $\rho_c$ of the as received and lightly oxidized stainless steel were sensitive to load cycling due to the wear of the very thin oxide layer. After 30,000 cycles, $\rho_c$ dropped up to 4 orders of magnitudes. This makes effective $\rho_c$ control practically impossible. The sensitivity to load cycling decreased with increasing oxidation temperature within the range of 300 – 500 °C. However, the $\rho_c$ values were too high for our application and not very responsive to oxidation temperature. Therefore, the oxidation of stainless steel alone does not seem to be suitable for $\rho_c$ control, and other methods of oxidation were experimented.

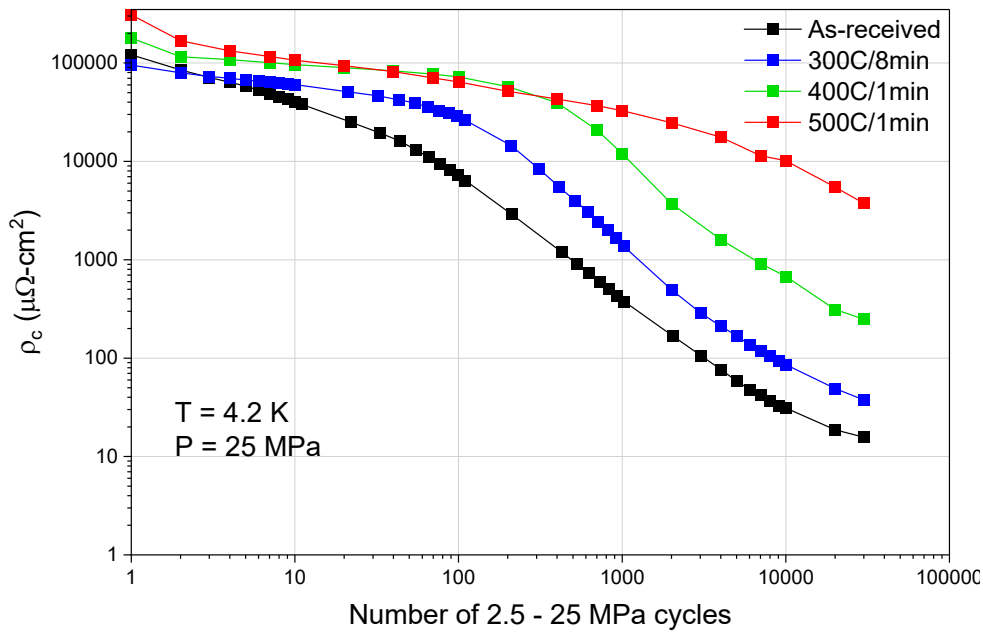

Figure 3. $\rho_c$ of REBCO/stainless steel/REBCO stack with stainless steel tapes oxidized at various temperatures and times. Except for the as-received one, the stainless steel tapes were etched by 37% HCl for 5 min at room temperature to remove the native oxides before oxidation.

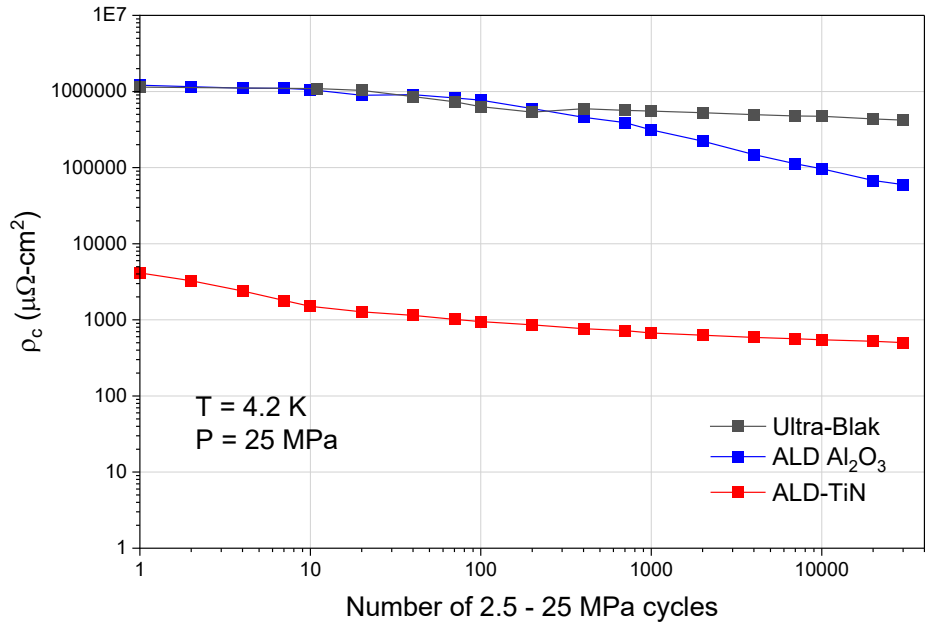

*Figure 4. $\rho_c$ vs. pressure cycles of 316L stainless steel treated by Ultra-Blak 407L, coated with TiN and Al$_2$O$_3$ by ALD.*

Figure 4 shows the results from stainless steel samples treated by Ultra-Blak for 15 min as well as the ones coated by TiN (300 layers) and Al$_2$O$_3$ (300 layers) by using atomic layer deposition (ALD). With these useful developments, the sensitivity to load cycling was much reduced. However, $\rho_c$ was still not very responsive to the change of process parameters, and scaling up these thin film growth techniques will be challenging. Additionally, it is difficult to obtain lower $\rho_c$ values using Ultra-Blak alone.

The load cycling sensitivity is likely due to wear at the contact asperities where the localized stress is high. Therefore, if the local stress is distributed more evenly by filling the voids with a material at the contacting interface as depicted in Figure 5 (a), the wear and load cycling sensitivity would be reduced. We applied paraffin wax and Stycast 1266 epoxy fillers respectively between REBCO and stainless steel. The paraffin was impregnated in vacuum at about 100 °C under a 25 MPa contact pressure, while the Stycast 1266 was applied between REBCO and stainless steel and cured in ambient under a 25 MPa contact pressure. Figure 5(b) shows their load cycling results. The $\rho_c$ values are comparable with their dry contacts counterparts even though both paraffin and epoxy are insulating, which suggests that the contact asperities were free of impregnating materials. The sample impregnated with epoxy showed much improved cycling sensitivity as predicted. But the paraffin impregnated sample did not show improvement, as it is too weak to distribute the loads at the asperities.

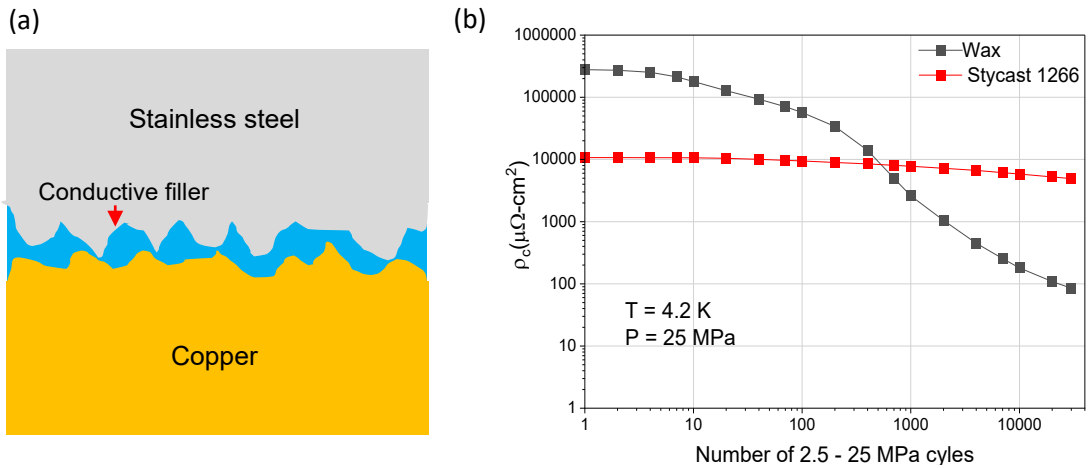

*Figure 5. (a) Schematic of a contacting interface filled with impregnating material. (b) $\rho_c$ of samples impregnated by paraffin wax and Stycast 1266 under 25 MPa contact pressure*

Encouraged by above results and with the goal of lowering $\rho_c$ to a range that is responsive to thermal oxidation treatment, we applied conductive fillers between the contacting surfaces using silver paint (DuPont 4929N) and silver epoxy (H20E by Epoxy Technology) respectively. The silver paint was cured in atmosphere for 16 hours, the silver epoxy was cured at 135 C for 15 min, both under 25 MPa pressure. The results show much improved load cycling robustness as shown in Figure 6. Meanwhile, the increase in the actual contact area by conductive fillers made $\rho_c$ considerably lower. For instance, $\rho_c$ with 500 °C oxidized stainless steel was 230,000 $\mu\Omega\text{-cm}^2$. With silver paint filler, $\rho_c$ was only 3,000 $\mu\Omega\text{-cm}^2$, a factor of 70 reduction. This way, a wide range of $\rho_c$ control can be achieved by changing temperature of thermal oxidation or using other oxidation methods as also shown in Figure 6.

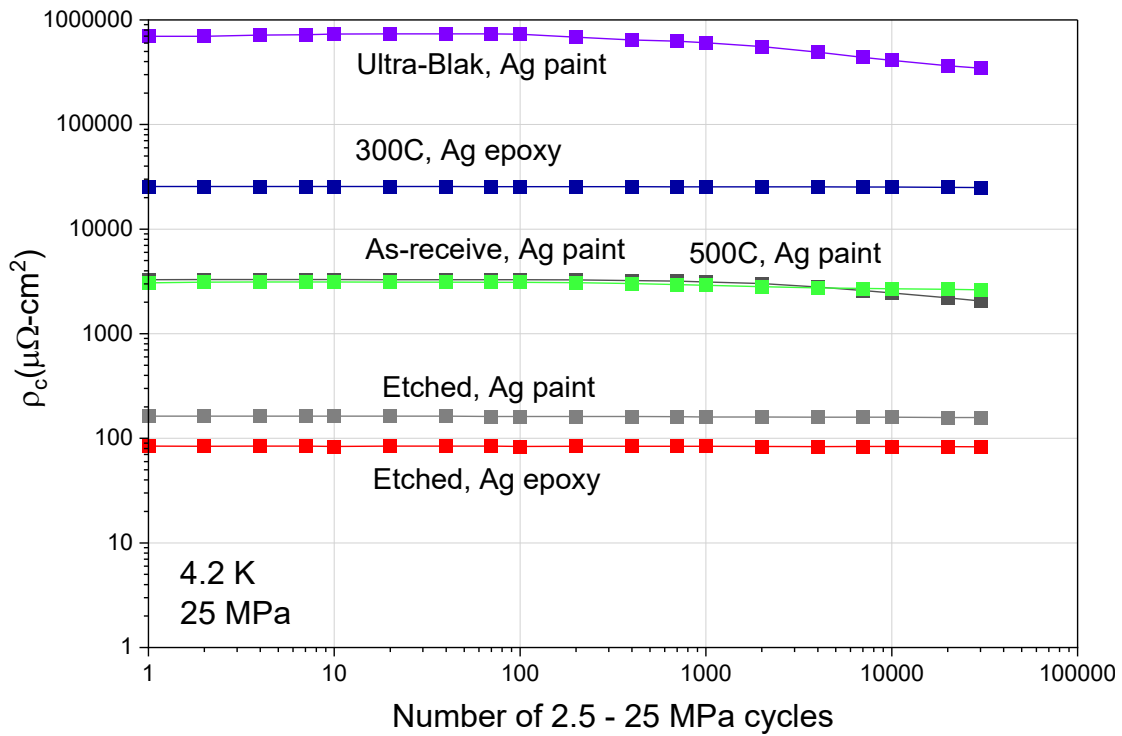

*Figure 6. $\rho_c$ versus load cycle for stacks of REBCO/stainless steel/REBCO where Ag paint (Dupont 4929N) and silver epoxy (H20E, Epoxy Technology) were used as fillers at contact interfaces.*

Another important experiment was to check the integrity of the oxide layers after a high transverse current in simulation of the radial current during a magnet quench. During a quench, the high radial transient current may irreversibly damage the oxide layer and change $\rho_c$. For this experiment, we used a capacitor bank to inject several 100 A current pulses across the 1 $cm^2$ contact at 4.2 K. Each pulse was several milliseconds long. For samples with $\rho_c$ lower than 1000 $\mu\Omega\text{-}cm^2$, after 5 pulses of such current injection, the effect was insignificant. For higher $\rho_c$ samples, however, as much as 50% reduction in $\rho_c$ was observed after 5 pulses.

To summarize this section, our preliminary research showed that the method of co-wind oxidation together with the use of a conductive filler seemed to allow for $\rho_c$ control without load cycling sensitivity. It should be noted that similar approach was used by K. Bouloukakis [19] and more recently by S. Park et al [24] where copper and graphite filled conductive epoxy were used respectively, and filler composition was adjusted to control $\rho_c$. The shortcoming of this method, however, is that turn-to-turn bonding by epoxy may introduce thermal stress which could lead to severe Ic degradation [27]. Furthermore, in the wet winding process, the excessive conductive fillers could cause undesirable short circuits. All these require significant development. Therefore, an alternative dry winding method was developed and will be described in the next section.

### 3.2 Control of $\rho_c$ by solder coating REBCO and oxidation of the stainless steel

Since the load cycling effect was attributed to the wear of the very thin layer of oxides on stainless steel surface against copper stabilizer layer of the REBCO tape, an intuitive method of mitigation is to coat REBCO tape with a thin layer of material that is significantly softer than the electroplated copper layer. For instance, the eutectic PbSn solder has a hardness of only 16 Hv at room temperature [28], much softer than that of electroplated Cu of 180 - 220 Hv [29]. Although hardness at cryogenic temperatures for PbSn solder is not found in the literature, it is speculated to be significantly lower than that of copper. Therefore, it is conceivable that solder coated REBCO results in reduction in load cycling sensitivity.

Figure. 7(a) shows a cross-section of a stainless steel tape between two solder-coated REBCO tapes. The average coating thickness was 2 – 3 µm which is sufficient to prevent copper from directly contacting the stainless steel. The surface morphology was characterized by laser confocal microscopy, and the comparison before and after solder coating is shown in Figure 7(b) and 7(c). The coating smoothed the original electroplated copper surface as shown by decrease of roughness Ra from 0.88 to 0.30 µm.

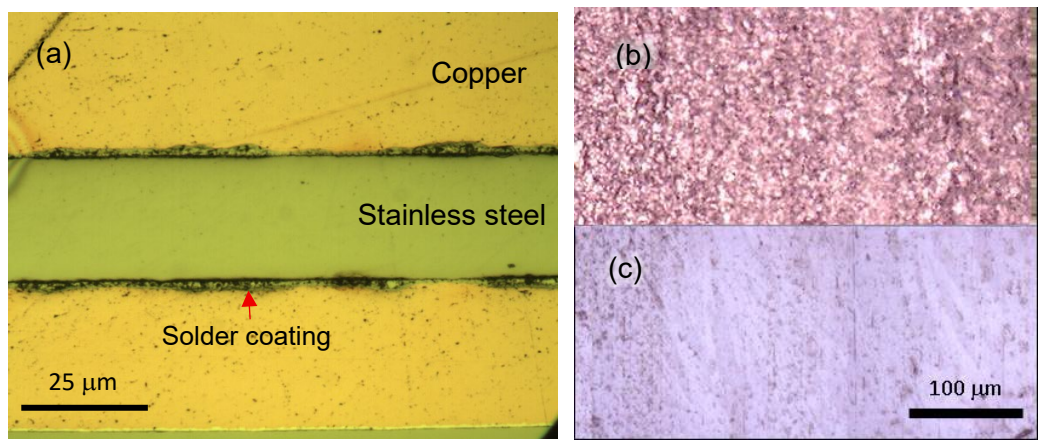

*Figure 7. (a) a cross-sectional micrograph of REBCO/Stainless steel/REBCO interfaces. REBCO is coated with PbSn solder which seems to smooth out the relatively rough surface of Cu stabilizer of REBCO. (b) laser confocal microscopy of the as-received REBCO surface and (c) solder coated REBCO surface.*

The solder coated REBCO were tested in REBCO/stainless steel/REBCO stacks (with no fillers). The results are shown in Figure 8. Evidently, by coating REBCO with solder, $\rho_c$ values were significantly lower than those shown in Figure 3, and the load cycling effect is significantly reduced. These made the process suitable for our test coil, where $\rho_c$ is to be controlled in the range of 1000 – 10000 µΩ-cm$^2$. This can be achieved by adjusting oxidation temperature of stainless steel co-wind tape in the range of 300 - 600 °C.

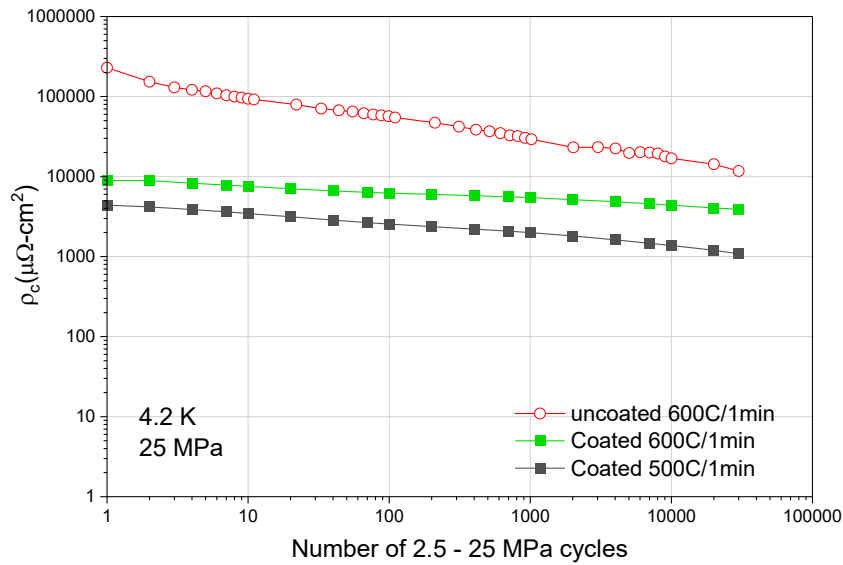

*Figure 8. The effect coating solder on REBCO. $\rho_c$ of REBCO/stainless steel/REBCO stack with solder-coated REBCO and an oxidized 25 µm thick stainless steel. Compared with non-coated case (data points of empty circle), initial $\rho_c$ is reduced by more than one order of magnitude.*

It was previously reported that $\rho_c$ between REBCO tapes is sensitive to temperatures [7]. With oxidized stainless steel co-wind, however, the temperature dependence of REBCO/stainless steel/REBCO is dominated by the properties of the oxide and behaves differently. Figure 9 is the $\rho_c$ temperature dependence measured under 25 MPa contact pressure, which shows that the temperature dependence is very weak. The implication of this result is that $\rho_c$ at 77 K can represent the results at 4.2 K. This allowed us to do most of $\rho_c$ control development at 77 K, which reduced costs and time considerably.

As a crucial step of the $\rho_c$ control, we oxidized stainless steel tapes in reel-to-reel fashion at different temperatures and measured $\rho_c$ of REBCO/stainless steel/REBCO, where the REBCO were solder coated. The oxidation temperature dependence of $\rho_c$ at 77 K and 10 MPa is shown in Figure 10. The $\rho_c$ control development was aimed to be 1000 and 5000 µΩ-cm² for test coils. Based on the results in Figure 10, oxidation temperatures of 415 °C and 500 °C were chosen to achieve 1000 and 5000 µΩ-cm² respectively. Their load cycling sensitivity was measured at 4.2 K (Figure 11), which is moderate and acceptable for future RI magnets.

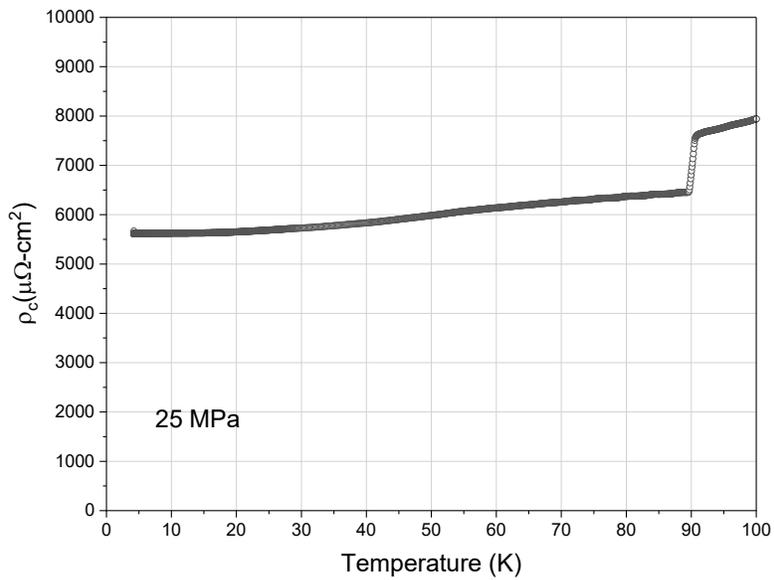

*Figure 9. $\rho_c$ versus temperature of a REBCO/stainless steel/REBCO. The stainless steel was in as-received condition with native oxide, silver epoxy (H20E) was applied between REBCO and stainless steel.*

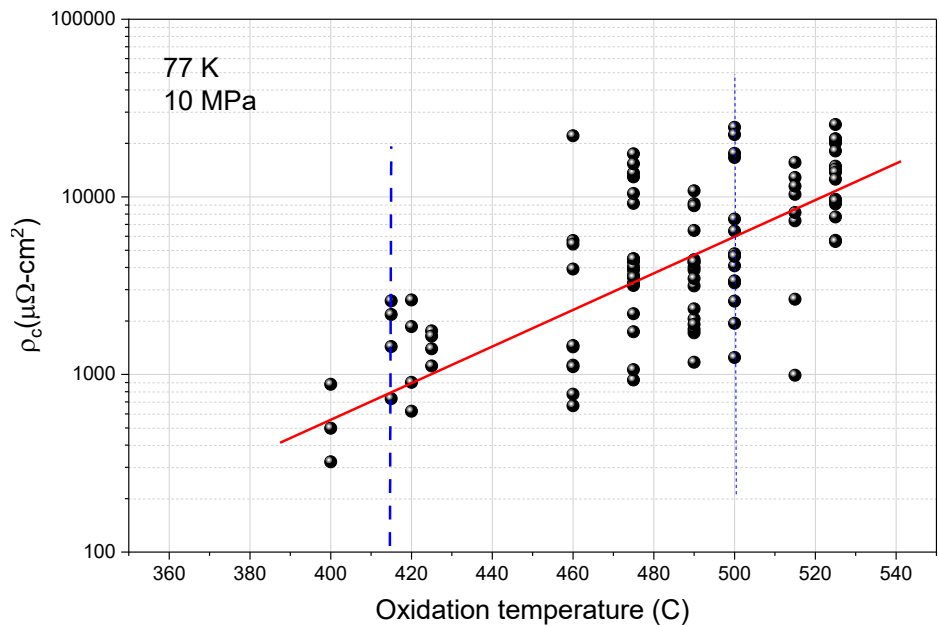

*Figure 10. $\rho_c$ of REBCO/stainless steel/REBCO at 77 K with 10 MPa contact pressure. REBCO samples were coated with solder, and the stainless steel were oxidized at different temperatures for 1 min. Due to the chimney effect of the vertical furnace, the actual temperature in the reel-to-reel oxidation system was not uniform and the maximum temperature which was at near the top of the furnace is typically 100 °C lower than the setpoints. The oxidation speed was about 1 m/min.*

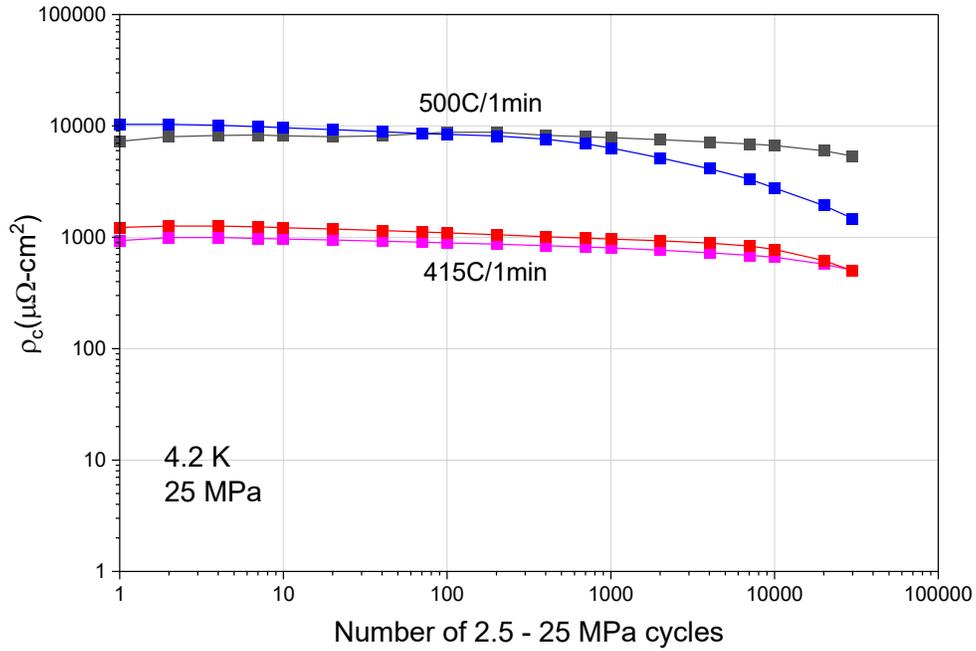

Figure 11.  $\rho_c$ versus transverse load cycles targeting 1000 and 5000 $\mu\Omega$-$cm^2$. Two samples were measured for each oxidation condition.

### 3.3 Long length oxidized stainless steel tapes

To use this $\rho_c$ control method described in section 3.2 in real magnet coils, it is necessary to ensure the consistency of $\rho_c$ along the length. It is also necessary to measure $\rho_c$ using the as-received stainless steel tapes because the thickness of native oxides on stainless steel surface is unknown and varies from batch to batch even by the same manufacturer. Figure 12 shows $\rho_c$ of samples cut from a 17 m long stainless steel tape oxidized at 400 °C for 1 min. The average $\rho_c$ value is 2000 $\mu\Omega$-$cm^2$ with a standard deviation of 388 $\mu\Omega$-$cm^2$, which is about 20% of the average value. Such a level of uncertainty must be considered in the quench calculation of RI coils.

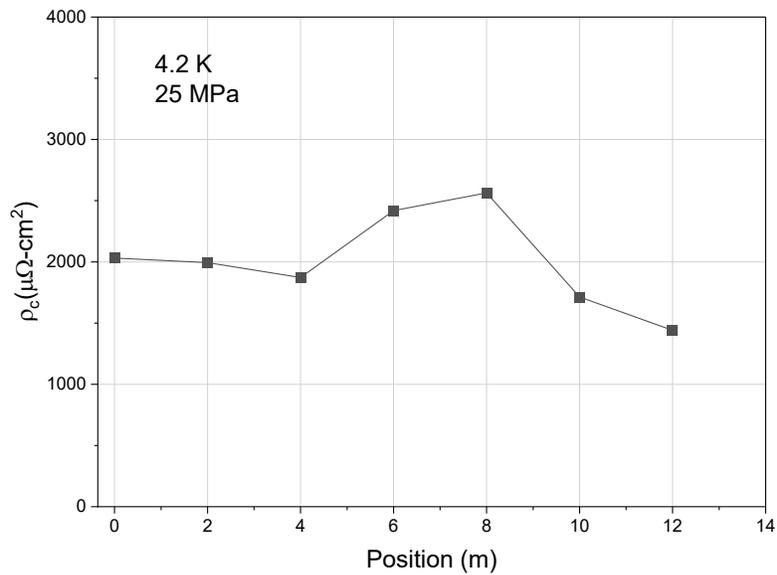

*Figure 12. Variation of $\rho_c$ along the length of stainless steel tape which was oxidized at 400 °C for 1 min. $\rho_c$ was measured at 77 K under 10 MPa contact pressure. The same pair of solder coated REBCO were used for above tests.*

For the PTC-6 test coil as described in the experimental section, developing $\rho_c$ of the 1000 $\mu\Omega$-cm$^2$ was needed. Based on Figure 10, 415 °C was chosen as the temperature for reel-to-reel oxidation. Subsequently, we measured $\rho_c$ of multiple piece lengths of stainless steel tapes using the same pair of REBCO tapes. Furthermore, we developed $\rho_c$ of the 5000 $\mu\Omega$-cm$^2$ for future RI coils of larger diameters. Both results are shown in Figure 13. Of all the stainless steel tapes of 1.2 and 6.4 km targeted for 1000 and 5000 $\mu\Omega$-cm$^2$ respectively, the variation in $\rho_c$ is on the order of 50%.

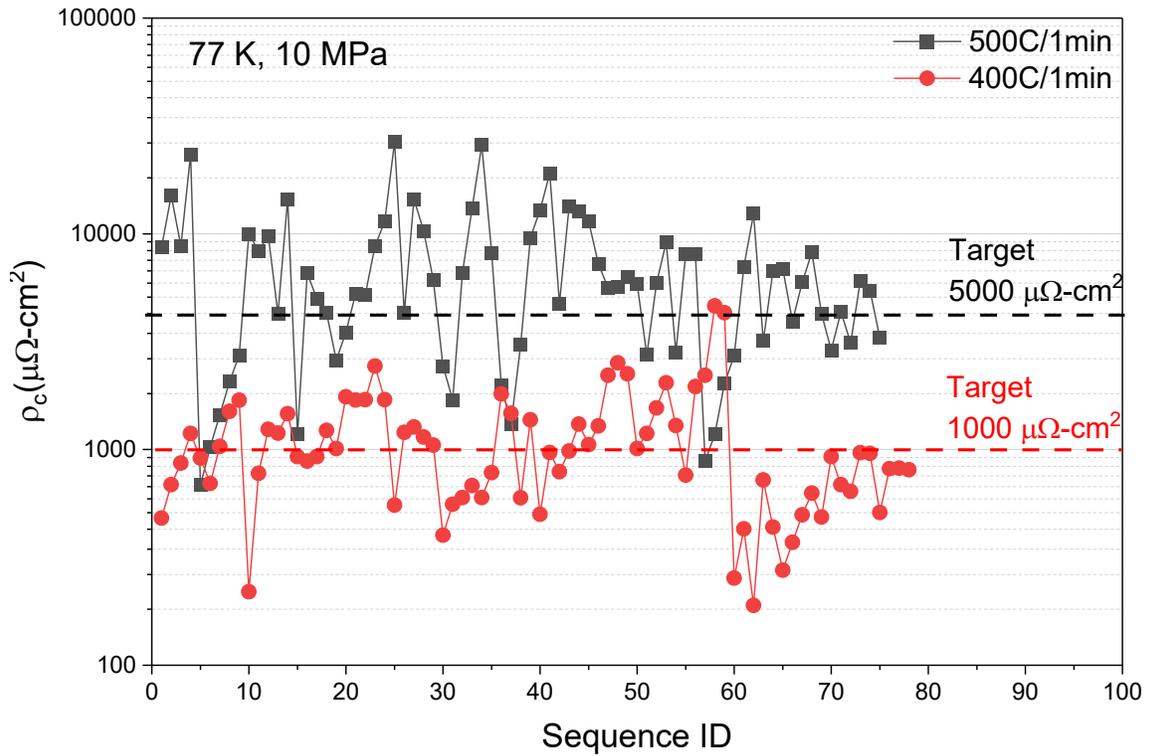

Figure 13. $\rho_c$ measured with an oxidized stainless steel sandwiched between two solder-coated REBCO tapes targeting (a) 1000 $\mu\Omega$-cm$^2$ for use in PTC-6 and (b) 5000 $\mu\Omega$-cm$^2$.

### 3.4 PTC-6 test coil

RI coil PTC-6 was fabricated and tested for major performance goals including $\rho_c$ [30]. The decay time constant after a sudden discharge at 5 A was measured; and $\rho_c$ was calculated using equations (1) and (2). The results are tabulated in Table 2 and plotted in Figure 14.

Table 2 $\rho_c$ of PTC-6 by sudden discharge at 5 A and 4.2 K

| No. | Event | $\tau$ (s) | Rc ($\Omega$) | $\rho_c$ ($\mu\Omega$-cm$^2$) |
|---|---|---|---|---|
| 1 | Before charging | 0.533 | 0.1553 | 817 |
| 2 | After HTS only 2 ramps to 240 A | 0.473 | 0.1771 | 930 |
| 3 | After 10 quench tests (one 240A, nine 210A) | 0.239 | 0.3467 | 1821 |
| 4 | After 2 thermal cycles | 0.214 | 0.3840 | 2017 |
| 5 | After 2 charging cycles to 230 A w/o LTS field | 0.184 | 0.4475 | 2350 |
| 6 | After 6 quench tests | 0.141 | 0.5843 | 3069 |
| 7 | After 7 more quench tests | 0.274 | 0.3003 | 1577 |
| 8 | After 500 cycles | 0.266 | 0.3091 | 1624 |
| 9 | After 500 cycles | 0.301 | 0.2729 | 1434 |
| 10 | At the end of coil test | 0.261 | 0.3145 | 1652 |

As shown in Table 2 and Figure 14, $\rho_c$ is in a range of 800 - 3000 $\mu\Omega\text{-cm}^2$. Two thermal cycles in event No. 4 were due to warming up the coil to room temperature for terminal repair. After the repair, the coil was cooled down to test in liquid nitrogen then warmed up followed by final cool down to 4.2 K . None of thermal cycling and the total of 1000 charge/discharge (210 – 100 A) cycles seems to change $\rho_c$ significantly. This result demonstrates that our control of contact resistivity is successful in both achieving the designed $\rho_c$ value and mitigation of load cycling sensitivity. Quenches induced by quench heaters, however, had significant impact on $\rho_c$, as shown at event 3, 6, and 7 in Figure 14.

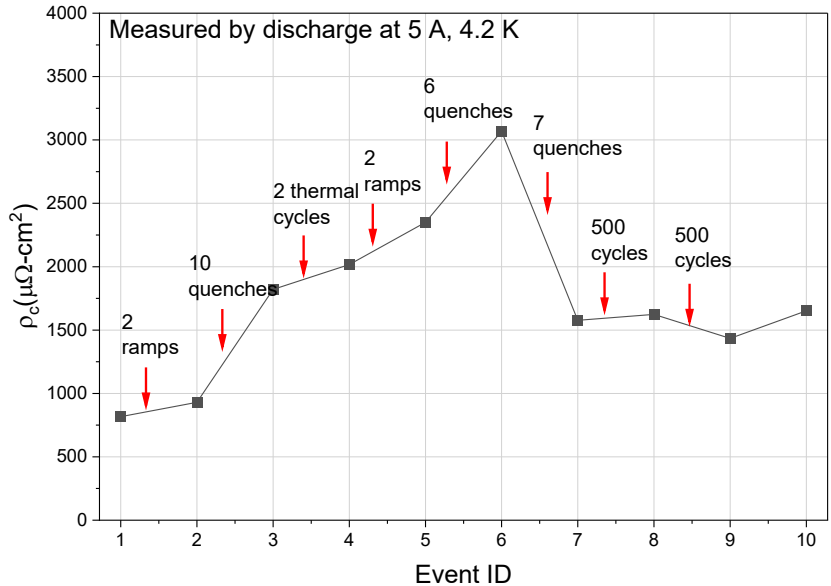

Figure 14. (a) PTC-6 $\rho_c$ measured after a sequence of events as listed in Table 2.

4. **Discussions**

Achieving reliable control of $\rho_c$ is extremely challenging. It is unlikely that turn-to-turn contact resistance can be adjusted once the coil is wound. It must be tuned in the coil fabrication process by changing either the contact area [31], or $\rho_c$ as in [23] and this work, or both [21],[32]. It also seems to be difficult to have a consistent $\rho_c$ over a long length as Figure 13 suggests. Similar variation in $\rho_c$ among multiple double-pancakes was reported to be about 50% [22]. In addition, $\rho_c$ varies with coil's radial stress, i.e. the contact pressure, which varies with radial position. Estimation of radial stress distribution is difficult due to large uncertainties in radial modulus of a dry-wound coil [33].

It is well known that real contacts are at asperities. Even with a seemingly perfect contact, the actual contact area is only a small fraction of the nominal contact area [34]. Since REBCO is vulnerable to further processing, the preferred way to control $\rho_c$ is by deposition of a highly resistive thin film on the co-wind tape. In this work, we showed that by combination of increasing contacting area with solder coating to decrease $\rho_c$ and oxidation of stainless steel co-wind to increase $\rho_c$, the $\rho_c$ can be adjusted in the range of 100 to 1000000 $\mu\Omega\text{-cm}^2$ with a relative uncertainty of about +/- 50%. The considerable variation in $\rho_c$ is understandable given that the variation in surface conditions of both REBCO and stainless steel, the actual contact area can vary significantly.

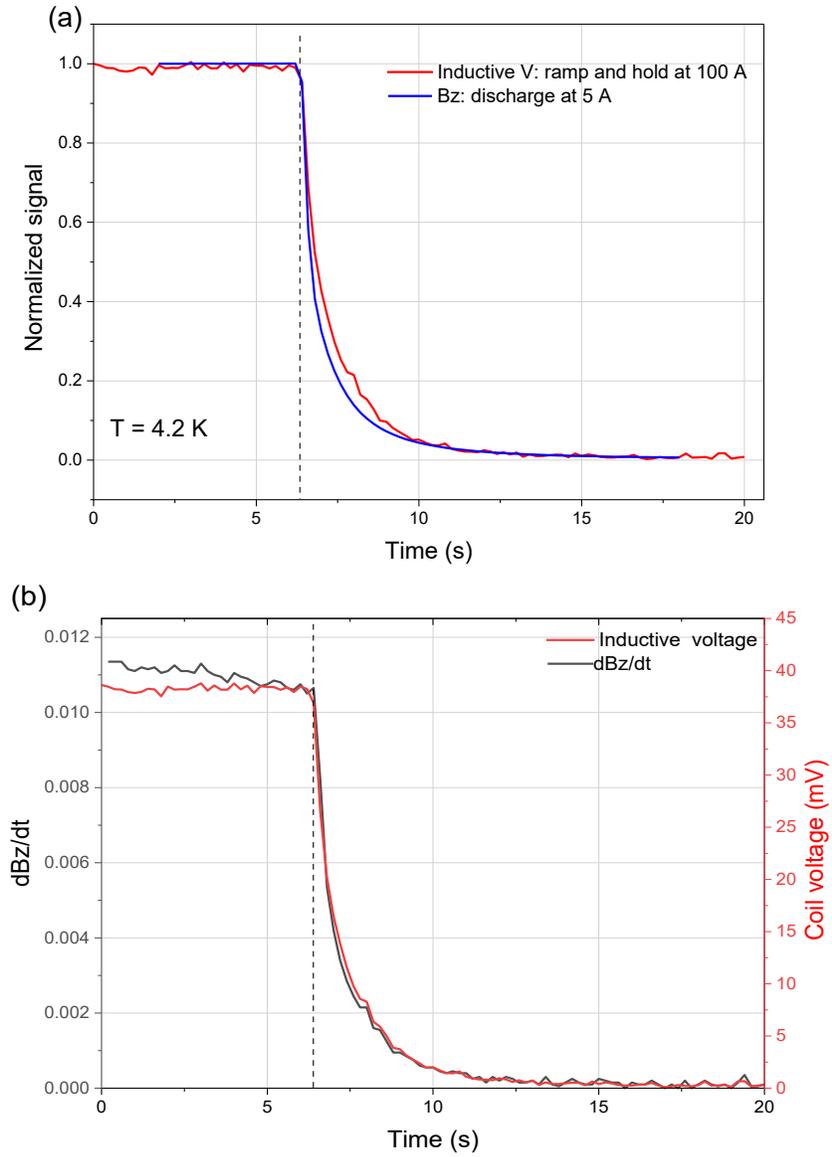

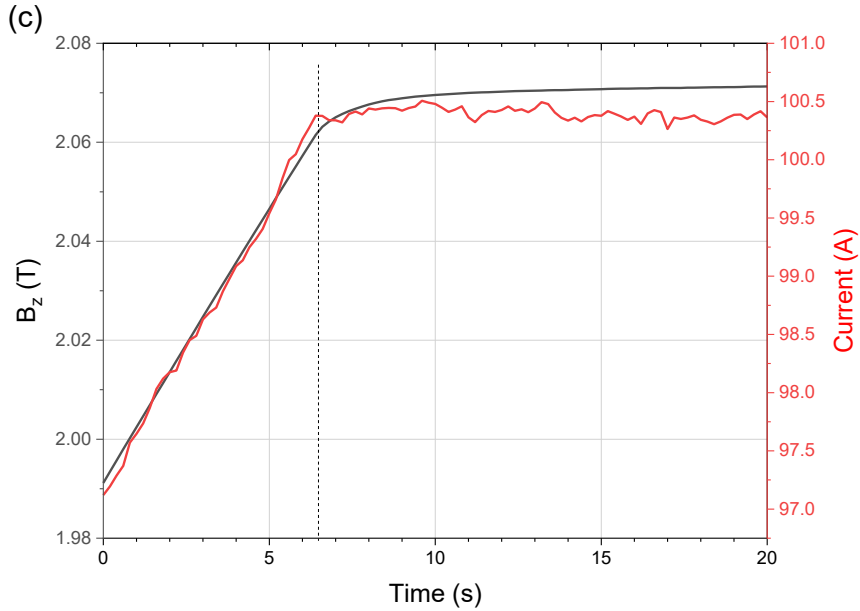

*Figure 15. (a) Comparison between inductive voltage decay after a 0.5 A/s ramp and a field Bz decay after a sudden discharge from 5 A. (b) The inductive voltage decay versus dBz/dt. (c) Current and field ramp. The vertical dashed lines indicate the onset of the decays.*

During testing of the RI coils, we found that the inductive voltage decayed to zero when a current ramp was held. The time-constant of this inductive voltage decay was comparable to the decay of central field Bz after a sudden discharge (Figure 15(a)). This can be explained by the fact that inductive voltage is proportional to dBz/dt. When Bz had a charging delay due to contact resistance, dBz/dt behaved as what is shown in Figure 15(b), which decays the same way as the inductive voltage. Figure 15(c) shows the current ramp and the Bz ramp with an apparent charging delay between 6.5 and 10 s. Therefore, the inductive voltage decay constant can be used for $\rho_c$ measurement. It should be noted that, although inductive voltages were measured from each double-pancake, the decay time constants are all the same as they all correspond to dBz/dt. For nested coils with different designed $\rho_c$ values, the measured time constant reflects an average of all the coils.

An important advantage of using inductive voltage method is that the measurement is done at the actual current and field conditions, so the effect of Lorentz stress on the contact is at the operating condition, whereas the sudden discharge tests are usually conducted at significantly lower current and fields to avoid coil damage. Y. S. Chae, et al [35] and J. T. Lee, et al [36] used more elaborate simulation of inductive voltage during current ramps to obtain $\rho_c$. Due to the uncertainty introduced by the magnet current ramp rate, time constant values are prone to uncertainties. Sudden discharge method, on the other hand, results in more reproducible time constant values, and it usually generates large turn-to-turn current which is closer to the situation of a magnet quench. Therefore, it is a good method to study the impact of $\rho_c$ on quench behavior and was used to generate data and presented in table 2 and Figure 14. It is noteworthy that the apparent contact resistivity increase with increasing discharging current was observed by S. Liu, et al [22].

**Conclusions**

We developed a reliable method for controlling contact resistance $\rho_c$ in resistively-insulated REBCO coils. In our initial development efforts, we demonstrated $\rho_c$ control by oxidation of stainless steel co-wind tapes. To mitigate the load cycling effect on $\rho_c$, silver paste and silver epoxy were applied at the interface between REBCO and stainless steel. While this wet winding method has its merit, it raises the concern that the epoxy may result in REBCO delamination. We found that in dry-wound coil the loading cycling effect is minimized by coating REBCO with a thin layer of PbSn solder. Both oxidation of stainless steel co-wind tapes and solder coating of REBCO tapes were done in a reel-to-reel fashion for long length application. REBCO and stainless steel tapes treated by these methods were used in a six double-pancake test coil. The coil test results showed that the average $\rho_c$ of the coil was in good agreement with the designed value of 1000 $\mu\Omega\text{-cm}^2$. A method of measuring coil's $\rho_c$ by the decay time of coil's inductive voltage is proposed.

**Acknowledgement**

We thank Dr. Eric Lochner for helping with atomic-layer-deposition of TiN and $Al_2O_3$ on stainless steel samples, Robert Goddard for confocal laser microscopy measurements, and Prof. Seungyong Hahn for providing the program for $\rho_c$ calculation. This work was performed at the National High Magnetic Field Laboratory, which is supported by National Science Foundation Cooperative Agreement No. DMR-1644779 and DMR-1839796, and the State of Florida.